# Fastest Distributed Consensus Averaging Problem on Perfect and Complete $n$-ary Tree networks

Saber Jafarizadeh
Department of Electrical Engineering
Sharif University of Technology, Azadi Ave, Tehran, Iran
Email: jafarizadeh@ee.sharif.edu

*Abstract*—Solving fastest distributed consensus averaging problem (i.e., finding weights on the edges to minimize the second-largest eigenvalue modulus of the weight matrix) over networks with different topologies is one of the primary areas of research in the field of sensor networks and one of the well known networks in this issue is tree network. Here in this work we present analytical solution for the problem of fastest distributed consensus averaging algorithm by means of stratification and semidefinite programming, for two particular types of tree networks, namely perfect and complete $n$-ary tree networks.

Our method in this paper is based on convexity of fastest distributed consensus averaging problem, and inductive comparing of the characteristic polynomials initiated by slackness conditions in order to find the optimal weights.

Also the optimal weights for the edges of certain types of branches such as perfect and complete $n$-ary tree branches are determined independently of rest of the network.

I. INTRODUCTION

Distributed computation in the context of computer science is a well studied field with an extensive body of literature (see, for example, [1] for early work), where some of its applications include distributed agreement, synchronization problems, [2] and load balancing in parallel computers [3,4].

A problem that has received renewed interest recently is distributed consensus averaging algorithms in sensor networks and one of main research directions is the computation of the optimal weights that yields the fastest convergence rate to the asymptotic solution [5, 6, 7] known as Fastest Distributed Consensus averaging Algorithm, which computes iteratively the global average of distributed data in a sensor network by using only local communications. Moreover algorithms for distributed consensus find applications in, e.g., multi-agent distributed coordination and flocking [8, 9, 10, 11], distributed data fusion in sensor networks [12, 13, 6], fastest mixing Markov chain problem [14], clustering [15, 16] gossip algorithms [17, 18], and distributed estimation and detection for decentralized sensor networks [19, 20, 21, 22, 23]. Recently in [24, 25, 26] the author has solved fastest distributed consensus averaging problem analytically for path, complete cored, symmetric and asymmetric star networks and in [27] the author has determined the optimal weights for edges of different types of branches of an arbitrary connected network and has proved that the obtained weights are independent of rest of the network.

One of basic and common types of networks is tree network also known as hierarchical network. Tree networks use a central hub called root node as the main communications router. One level down from the root node in the hierarchy is a central hub. This lower level then forms a Star network. Therefore tree network can be considered a hybrid of both the Star and Peer to Peer networking topologies. The tree network topology is ideal when the sensors are located in groups, with each group occupying a relatively small physical region. An example is a university campus in which each building has its own star network, and all the central computers are linked in a campus-wide system. In a tree network, a link failure in one of the star networks will isolate the sensors linked to the central node of that star network, but only those sensors will be isolated. All the other sensors will continue to function normally, except that they will not be able to communicate with the isolated sensors.

Here in this work, we have provided analytical solution for fastest distributed consensus averaging problem over two particular types of tree networks, namely perfect $n$-ary tree and complete $n$-ary tree networks, by means of stratification and semidefinite programming. Our method in this paper is based on convexity of fastest distributed consensus averaging problem, and inductive comparing of the characteristic polynomials initiated by slackness conditions in order to find the optimal weights. Also it is indicated that the obtained optimal weights hold for perfect $n$-ary tree branch and complete $n$-ary tree branch independent of rest of the network where these branches are obtained from connecting a perfect $n$-ary tree network and complete $n$-ary tree network to an arbitrary connected network via a bridge, respectively.

The organization of the paper is as follows. Section II is an overview of the materials used in the development of the paper, including relevant concepts from distributed consensus averaging algorithm, graph symmetry and semidefinite programming. Section III contains the main results of the paper where perfect $n$-ary tree network and complete $n$-ary tree network along with perfect and complete $n$-ary tree branches are introduced together with the corresponding evaluated *SLEM* and obtained optimal weights. Sections IV is devoted to proof of main results of paper for perfect $n$-ary tree network and section V concludes the paper.

## II. PRELIMINARIES

This section introduces the notation used in the paper and reviews relevant concepts from distributed consensus averaging algorithm, graph symmetry and semidefinite programming.

### A. Distributed Consensus

We consider a network $\mathcal{N}$ with the associated graph $\mathcal{G} = (\mathcal{V}, \mathcal{E})$ consisting of a set of nodes $\mathcal{V}$ and a set of edges $\mathcal{E}$ where each edge $\{i, j\} \in \mathcal{E}$ is an unordered pair of distinct nodes.

Each node $i$ holds an initial scalar value $x_i(0) \in \mathbf{R}$, and $x^T(0) = (x_1(0), \dots, x_n(0))$ denotes the vector of initial node values on the network. Within the network two nodes can communicate with each other, if and only if they are neighbors.

The main purpose of distributed consensus averaging is to compute the average of the initial values, $(1/n) \sum_{i=1}^{n} x_i(0)$ via a distributed algorithm, in which the nodes only communicate with their neighbors.

In this work, we consider distributed linear iterations, which have the form

$$x_i(t+1) = W_{ii} x_i(t) + \sum_{j \neq i} W_{ij} x_j(t), \quad i = 1, \dots, n$$

where $t = 0, 1, 2, \dots$ is the discrete time index and $W_{ij}$ is the weight on $x_j$ at node $i$ and the weight matrix have the same sparsity pattern as the adjacency matrix of the network's associated graph or $W_{ij} = 0$ if $\{i, j\} \notin \mathcal{E}$, this iteration can be written in vector form as

$$x(t+1) = W x(t)$$

The linear iteration (1) implies that $x(t) = W^t x(0)$ for $= 0, 1, 2, \dots$. We want to choose the weight matrix $W$ so that for any initial value $x(0)$, $x(t)$ converges to the average vector $\bar{x} = (\mathbf{1}^T x(0)/n)\mathbf{1} = (\mathbf{1}\mathbf{1}^T/n)x(0)$ i.e.

$$\lim_{t \to \infty} x(t) = \lim_{t \to \infty} W^t x(0) = \frac{\mathbf{1}\mathbf{1}^T}{n} x(0)$$

(Here $\mathbf{1}$ denotes the column vector with all coefficients one). This is equivalent to the matrix equation

$$\lim_{t \to \infty} W^t = \frac{\mathbf{1}\mathbf{1}^T}{n} \tag{1}$$

Assuming (1) holds, the *convergence factor* can be defined as

$$r(W) = \sup \frac{\|x(t+1) - \bar{x}\|_2}{\|x(t) - \bar{x}\|_2}$$

where $\|\cdot\|_2$ denotes the spectral norm, or maximum singular value. The FDC problem in terms of the convergence factor can be expressed as the following optimization problem:

$$\begin{aligned}
&\min_{W} \quad r(W) \\
&s.t. \quad \lim_{t \to \infty} W^t = \mathbf{1}\mathbf{1}^T/n, \\
&\quad \forall \{i,j\} \notin \mathcal{E}: W_{ij} = 0
\end{aligned} \quad (2)$$

where $W$ is the optimization variable, and the network is the problem data.

The FDC problem (2) is closely related to the problem of finding the fastest mixing Markov chain on a graph [14]; the only difference in the two problem formulations is that in the FDC problem, the weights can be (and the optimal ones often are) negative, hence faster convergence could be achieved compared with the fastest mixing Markov chain on the same graph.

In [5] it has been shown that the necessary and sufficient conditions for the matrix equation (1) to hold is that one is a simple eigenvalue of $W$ associated with the eigenvector $\mathbf{1}$, and all other eigenvalues are strictly less that one in magnitude. Moreover in [5] FDC problem has been formulated as the following minimization problem

$$\begin{aligned}
&\min_{W} \quad \max(\lambda_2, -\lambda_n) \\
&s.t. \quad W = W^T, W\mathbf{1} = \mathbf{1} \\
&\quad \forall \{i,j\} \notin \mathcal{E}: W_{ij} = 0
\end{aligned}$$

where $1 = \lambda_1 \geq \lambda_2 \geq \cdots \geq \lambda_n \geq -1$ are eigenvalues of $W$ arranged in decreasing order and $\max(\lambda_2, -\lambda_n)$ is the *Second Largest Eigenvalue Modulus* (*SLEM*) of $W$, and the main problem can be derived in the semidefinite programming form as [5]:

$$\begin{aligned}
&\min_{W} \quad s \\
&s.t. \quad -sI \preccurlyeq W - \mathbf{1}\mathbf{1}^T/n \preccurlyeq sI \\
&\quad W = W^T, W\mathbf{1} = \mathbf{1} \\
&\quad \forall \{i,j\} \notin \mathcal{E}: W_{ij} = 0
\end{aligned} \quad (3)$$

We refer to problem (3) as the Fastest Distributed Consensus (FDC) averaging problem.

## B. Symmetry of Graphs

An automorphism of a graph $\mathcal{G} = (\mathcal{V}, \mathcal{E})$ is a permutation $\sigma$ of $\mathcal{V}$ such that $\{i,j\} \in \mathcal{E}$ if and only if $\{\sigma(i), \sigma(j)\} \in \mathcal{E}$, the set of all such permutations, with composition as the group operation, is called the automorphism group of the graph and denoted by $Aut(\mathcal{G})$. For a vertex $i \in \mathcal{V}$, the set of all images $\sigma(i)$, as $\sigma$ varies through a subgroup $G \subseteq Aut(\mathcal{G})$, is called the orbit of $i$ under the action of $G$. The vertex set $\mathcal{V}$ can be written as disjoint union of distinct orbits. In [28], it has been shown that the weights on the edges within an orbit must be the same.

## C. Semidefinite Programming

SDP is a particular type of convex optimization problem [29]. An SDP problem requires minimizing a linear function subject to a linear matrix inequality (LMI) constraint [30]:

$$\min \quad \rho = c^T x,$$

$$s.t. \quad F(x) \geq 0$$

where $c$ is a given vector, $x^T = (x_1, \dots, x_n)$, and $F(x) = F_0 + \sum_i x_i F_i$, for some fixed Hermitian matrices $F_i$. The inequality sign in $F(x) \geq 0$ means that $F(x)$ is positive semidefinite.

This problem is called the primal problem. Vectors $x$ whose components are the variables of the problem and satisfy the constraint $F(x) \geq 0$ are called primal feasible points, and if they satisfy $F(x) \geq 0$, they are called strictly feasible points. The minimal objective value $c^T x$ is by convention denoted by $\rho^*$ and is called the primal optimal value.

Due to the convexity of the set of feasible points, SDP has a nice duality structure, with the associated dual program being:

$$\max \quad -Tr[F_0 Z]$$

$$s.t. \quad Z \geq 0$$

$$Tr[F_i Z] = c_i$$

Here the variable is the real symmetric (or Hermitian) positive matrix $Z$, and the data $c, F_i$ are the same as in the primal problem. Correspondingly, matrix $Z$ satisfying the constraints is called dual feasible (or strictly dual feasible if $Z > 0$). The maximal objective value of $-Tr[F_0 Z]$, i.e. the dual optimal value is denoted by $d^*$.

The objective value of a primal (dual) feasible point is an upper (lower) bound on $\rho^*(d^*)$. The main reason why one is interested in the dual problem is that one can prove that $d^* \leq \rho^*$, and under relatively mild assumptions, we can have $\rho^* = d^*$. If the equality holds, one can prove the following optimality condition on $x$.

A primal feasible $x$ and a dual feasible $Z$ are optimal, which is denoted by $\hat{x}$ and $\hat{Z}$, if and only if

$$F(\hat{x})\hat{Z} = \hat{Z}F(\hat{x}) = 0. \tag{4}$$

This latter condition is called the complementary slackness condition.

In one way or another, numerical methods for solving SDP problems always exploit the inequality $d \leq d^* \leq \rho^* \leq \rho$, where $d$ and $\rho$ are the objective values for any dual feasible point and primal feasible point, respectively. The difference

$$\rho^* - d^* = c^T x + Tr[F_0 Z] = Tr[F(x)Z] \geq 0$$

is called the duality gap. If the equality $d^* = \rho^*$ holds, i.e. the optimal duality gap is zero, and then we say that strong duality holds.

### III. MAIN RESULTS

This section presents the main results of the paper. Here we have introduced perfect $n$-ary tree network and complete $n$-ary tree network along with perfect and complete $n$-ary tree branches together with the corresponding evaluated *SLEM* and optimal weights, proofs and more detailed discussion are deferred to Sections IV and V.

*A. Perfect n-ary tree Network*

Perfect $n$-ary tree graph is a $n$-ary tree graph with all leaf nodes at same depth and all internal nodes have degree $n$, where a leaf node in a tree is a node without any children and an internal node is the one that is not a leaf node and depth of a node is the distance from the node to the root (the only node without parent) of the tree. A perfect $n$-ary tree graph is depicted in Fig.1 for $n = 3, m = 3$.

In section IV we have proved that the optimal weights for the edges of a perfect $n$-ary tree network equals $1/(n + 1)$, except for the edges (weighted by $w_1$ in Fig.1.) connecting the nodes on level 1 to the root node, where their optimal weights equal $2/(n + 2)$, also *SLEM* of perfect $n$-ary tree network for different values of $m$ and $n$ is presented in table.1.

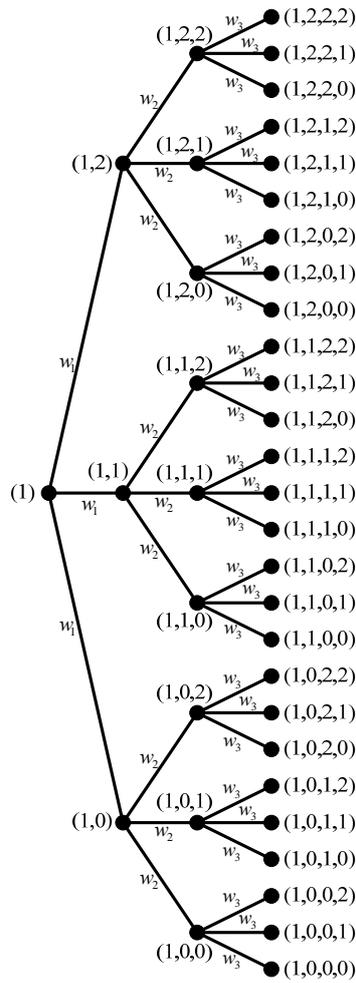

Fig.1. Perfect $n$-ary tree network for $n = 3, m = 3$.

| $(m,n)$ | SLEM | $(m,n)$ | SLEM |
| --- | --- | --- | --- |
| (3,2) | 0.958624546348495 | (3,4) | 0.989021583365371 |
| (5,2) | 0.992584711824087 | (5,4) | 0.999366349695859 |
| (10,2) | 0.999802655646275 | (10,4) | 0.999999386911883 |
| (3,3) | 0.980244608668151 | (3,5) | 0.993269038820414 |
| (5,3) | 0.998111423381200 | (5,5) | 0.999743114540051 |
| (10,3) | 0.999992471202028 | (10,5) | 0.999999918079809 |

Table.1. SLEM of perfect $n$-ary tree network for different values of $m$ and $n$

## B. Complete $n$-ary tree Network

Complete $n$-ary tree network is a more generalized form of perfect $n$-ary tree network. Complete $n$-ary tree network is a perfect $n$-ary tree network where all internal nodes at depth $i$ have degree $n_i$.

Here we consider a complete $n$-ary tree network with maximum depth $m$ and the undirected associated connectivity graph $\mathcal{G} = (\mathcal{V}, \mathcal{E})$. We denote the set of nodes on $i$-th depth of perfect $n$-ary tree graph by $\{(1, k_1, k_2, \ldots, k_i)\}$ where $k_j$ varies from 0 to $n_i - 1$ for $j = 1, \ldots, i$ with $i = 1, \ldots, m$ and we indicate the root of tree graph by $(1)$ (see Fig.2 for $m = 3, n_1 = 3, n_2 = 2, n_3 = 3$).

Automorphism of complete $n$-ary tree graph is $S_{N_i}$ permutation of nodes of same $i$-th depth for $i = 1, \ldots, m$, where $N_i = \prod_{j=1}^{i} n_j$ hence according to subsection II-B it has $m$ class of edge orbits and it suffices to consider just $m$ weights $w_1, w_2, \ldots, w_m$ (as labeled in Fig. 2. for $m = 3, n_1 = 3, n_2 = 2, n_3 = 3$).

Using the same procedure as in section V, namely stratification and semidefinite programming we can state that the optimal weights for the edges at depth $i$ of a complete $n$-ary tree network equals $1/(n_i + 1)$ for $i = 2, \ldots, m$, except for the edges at depth 1 (weighted by $w_1$ in Fig.2.) connecting the nodes on level 1 to the root node, where their optimal weights equal $2/(n_1 + 2)$, also *SLEM* of complete $n$-ary tree network for different values of $m$ and $n_i$ is presented in table.2.

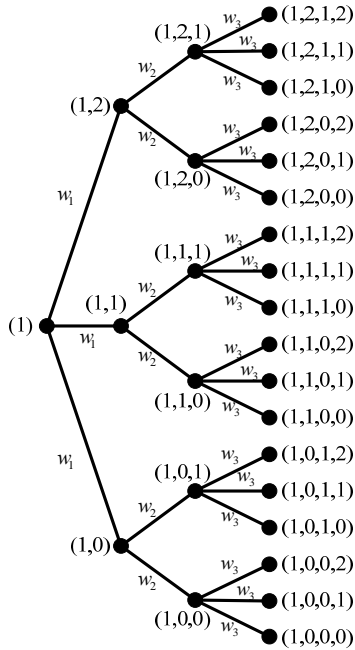

Fig.2. Complete $n$-ary tree network for $m = 3, n_1 = 3, n_2 = 2, n_3 = 3$.

| $(m, n_1, ..., n_m)$ | SLEM | $(m, n_1, ..., n_m)$ | SLEM |
|---|---|---|---|
| (2,2,3) | 0.911437827766148 | (3,2,3,5) | 0.984893852663005 |
| (2,2,4) | 0.930073525436772 | (3,3,4,5) | 0.989652912530167 |
| (2,3,3) | 0.924499799839840 | (4,2,3,4,5) | 0.996489302651042 |
| (2,3,4) | 0.940312423743285 | (4,3,3,4,5) | 0.996901389096411 |
| (3,2,3,4) | 0.981884577981674 | (4,2,4,4,5) | 0.997221160514972 |
| (3,3,3,4) | 0.984154938310847 | (4,2,3,4,6) | 0.996977318781077 |
| (3,2,4,4) | 0.985761127165554 | (4,3,4,5,6) | 0.998300241753577 |

Table.2. *SLEM* of complete $n$-ary tree network for different values of $m$ and $n_1, ..., n_m$.

## C. Perfect and Complete n-ary tree Branches

Perfect and complete $n$-ary tree branches consist of a perfect and complete $n$-ary tree network connected to an arbitrary network by a bridge, respectively. In Fig.3. a complete $n$-ary tree branch for $m = 3, n_1 = 3, n_2 = 2, n_3 = 3$ is depicted.

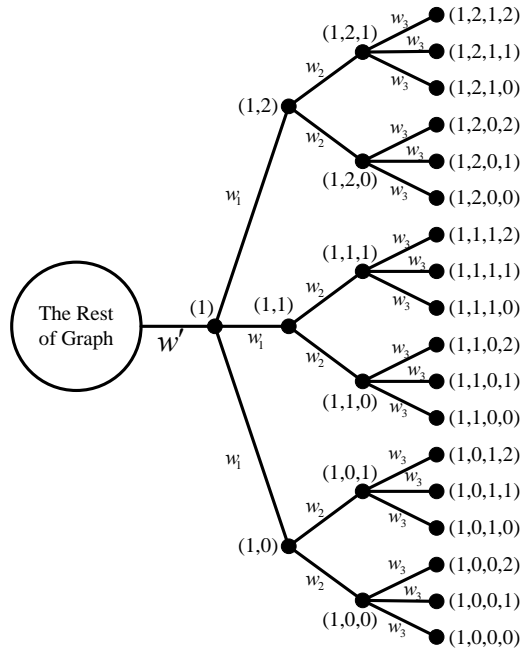

Fig.3. Complete $n$-ary tree branch for $m = 3, n_1 = 3, n_2 = 2, n_3 = 3$.

Using the same inductive procedure as in [27], we can state that the optimal weights for perfect and complete $n$-ary tree branches are the same as the optimal weights obtained in sections IV and V for perfect and complete $n$-ary tree networks respectively except for the bridge (weighted by $w'$ in Fig.3.) which is connecting perfect and complete $n$-ary tree networks to the arbitrary network. Specifying optimal weight of bridge and *SLEM* of the network requires more knowledge about full topology of network.

## IV. Proofs of Main Results

In this section solution of fastest distributed consensus averaging problem and determination of optimal weights for perfect $n$-ary tree network introduced in section III is presented.

Here we consider a perfect $n$-ary tree network with maximum depth $m$ and the undirected associated connectivity graph $\mathcal{G} = (\mathcal{V}, \mathcal{E})$. We denote the set of nodes on $i$-th depth of perfect $n$-ary tree graph by $\{(1, k_1, k_2, \ldots, k_i)\}$ where $k_j$ varies from 0 to $n-1$ for $j = 1, \ldots, i$ with $i = 1, \ldots, m$ and we indicate the root of tree graph by $(1)$ (see Fig.1 for $n = 3, m = 3$).

Automorphism of perfect $n$-ary tree graph is $S_{n^i}$ permutation of nodes of same $i$-th depth for $i = 1, \ldots, m$, hence according to subsection II-B it has $m$ class of edge orbits and it suffices to consider just $m$ weights $w_1, w_2, \ldots, w_m$ (as labeled in Fig. 1. for $n = 3, m = 3$) and consequently the weight matrix for the network can be defined as

$$W_{i,j} = \begin{cases} 1 - nw_1 & \text{for } i = j = (1), \\ 1 - w_m & \text{for } i = j = (1, k_1, \ldots, k_m), \ k_1, \ldots, k_m = 0, \ldots, n-1, \\ 1 - nw_{\mu+1} - w_\mu & \text{for } i = j = (1, k_1, \ldots, k_\mu), \ k_1, \ldots, k_\mu = 0, \ldots, n-1, \ \mu = 1, \ldots, m-1, \\ w_\mu & \text{for } i = (1, k_1, \ldots, k_{\mu-1}), \ j = (1, k_1, \ldots, k_{\mu-1}, k_\mu), \ k_1, \ldots, k_\mu = 0, \ldots, n-1, \ \mu = 1, \ldots, m \end{cases}$$

We associate with the node $(i)$, the $|\mathcal{V}| \times 1$ column vector $e_i \in \mathbf{R}^{|\mathcal{V}|}$ with 1 in the $i$-th position, and zero elsewhere where $|\mathcal{V}| = \frac{n^{m+1}-1}{n-1}$ is the total number of nodes on graph. Introducing the new basis $\varphi_1 = e_1$ and

$$\varphi_{1,k_1,\ldots,k_\mu} = \frac{1}{\sqrt{n^\mu}} \sum_{i_1,\ldots,i_\mu=0}^{n-1} \omega^{\sum_{j=1}^{\mu} k_j i_j} e_{1,i_1,\ldots,i_\mu} \quad \text{for } \mu = 1, \ldots, m$$

with $\omega = e^{j\frac{2\pi}{n}}$, the weight matrix $W$ for perfect $n$-ary tree network in the new basis takes the block diagonal form with diagonal blocks $W_1, W_2, \ldots, W_m$ where $W_1$ is defined as:

$$W_1 = \begin{bmatrix} 1 - nw_1 & \sqrt{n}w_1 & 0 & & \\ \sqrt{n}w_1 & 1 - w_1 - nw_2 & \sqrt{n}w_2 & 0 & \\ 0 & \sqrt{n}w_2 & 1 - w_2 - nw_3 & \ddots & 0 \\ & 0 & & \ddots & \ddots & \sqrt{n}w_m \\ & & & & \sqrt{n}w_m & 1 - w_m \end{bmatrix} \quad (5\text{-a})$$

and $W_i$ for $i = 2, ..., m$ is obtained from cutting first row and column of $W_{i-1}$ and consequently $W_2$ is:

$$W_2 = \begin{bmatrix} 1 - w_1 - nw_2 & \sqrt{n}w_2 & 0 & & \\ \sqrt{n}w_2 & 1 - w_2 - nw_3 & \sqrt{n}w_3 & 0 & \\ 0 & \sqrt{n}w_3 & 1 - w_3 - nw_4 & \ddots & 0 \\ & 0 & \ddots & \ddots & \sqrt{n}w_m \\ & & & \sqrt{n}w_m & 1 - w_m \end{bmatrix} \quad (5\text{-b})$$

Considering the fact that $W_2$ is a submatrix of $W_1$ and using *Cauchy Interlacing Theorem*,

*Theorem 1* (*Cauchy Interlacing Theorem*) [31]:

Let $A$ and $B$ be $n \times n$ and $m \times m$ matrices, where $m \leq n$, $B$ is called a compression of $A$ if there exists an orthogonal projection $P$ onto a subspace of dimension $m$ such that $PAP = B$. The Cauchy interlacing theorem states that If the eigenvalues of $A$ are $\lambda_1(A) \leq \cdots \leq \lambda_n(A)$, and those of $B$ are $\lambda_1(B) \leq \cdots \leq \lambda_m(B)$, then for all $j$,

$$\lambda_j(A) \leq \lambda_j(B) \leq \lambda_{n-m+j}(A)$$

Notice that, when $n - m = 1$, we have

$$\lambda_j(A) \leq \lambda_j(B) \leq \lambda_{j+1}(A)$$

we can state the following corollary for the eigenvalues of $W_1$ and $W_2$.

In the case of $n = 1$, after stratification the weight matrix $W$ does not include $W_2$ and $W$ reduces to $W_1$ and consequently Cauchy interlacing theorem will not be true thus the followings are true for $n \geq 2$.

*Corollary 1*,

If we consider $W_1$ and $W_2$ given in (5) then theorem 1 implies the following relations between the eigenvalues of $W_1$ and $W_2$

$$\lambda_{m_1+1}(W_1) \leq \lambda_{m_1}(W_2) \leq \cdots \leq \lambda_2(W_2) \leq \lambda_2(W_1) \leq \lambda_1(W_2) \leq \lambda_1(W_1) = 1$$

considering the fact that $W_i$ for $i = 3, ..., m$ is a sub matrix of $W_2$ and applying *Cauchy Interlacing Theorem* for $W_i$ for $i = 2, ..., m$, we can conclude that the eigenvalues $\lambda_2(W)$ and $\lambda_{|V|}(W)$ are amongst the eigenvalues of $W_2$ and $W_1$, respectively.

Based on the corollary 1 and subsection II-A, one can express FDC problem for perfect $n$-ary tree network in the form of semidefinite programming as:

$$\begin{aligned} \min \quad & s \\ s.t. \quad & W_2 \leq sI \\ & -sI \leq W_1 - \boldsymbol{vv}^T \end{aligned} \quad (6)$$

where $\boldsymbol{v}$ is a $(m + 1) \times 1$ column vector defined as:

$$v_i = \frac{\sqrt{n}-1}{n^{\frac{m+1}{2}}} \times n^{\frac{i-1}{2}} \quad \text{for} \quad i = 1, \ldots, m+1$$

which is eigenvector of $W$ corresponding to the eigenvalue one. The weight matrix $W$ can be written as

$$W_1 = I_{m+1} - \sum_{i=1}^{m} w_i\, \alpha_i \alpha_i^T \tag{7-a}$$

$$W_2 = I_m - \sum_{i=1}^{m} w_i\, \alpha_i' \alpha_i'^T \tag{7-b}$$

where $\alpha_i$ and $\alpha_i'$ are $(m+1) \times 1$ and $m \times 1$ column vectors defined as:

$$\alpha_i(j) = \begin{cases} \sqrt{n} & j = i \\ -1 & j = i+1 \\ 0 & \text{Otherwise} \end{cases} \quad \text{for} \quad i = 1, \ldots, m$$

$$\alpha_1' = \begin{cases} 1 & j = 1 \\ 0 & \text{Otherwise} \end{cases}$$

$$\alpha_i' = \begin{cases} \sqrt{n} & j = i-1 \\ -1 & j = i \\ 0 & \text{Otherwise} \end{cases} \quad \text{for} \quad i = 2, \ldots, m,$$

respectively. In order to formulate problem (6) in the form of standard semidefinite programming described in section II-C, we define $F_i, c_i$ and $x$ as below:

$$F_0 = \begin{bmatrix} I_{m+1} - vv^T & 0 \\ 0 & -I_m \end{bmatrix}$$

$$F_i = \begin{bmatrix} -\alpha_i \alpha_i^T & \\ & \alpha_i' \alpha_i'^T \end{bmatrix} \quad \text{for} \quad i = 1, \ldots, m,$$

$$F_{m+1} = I_{2m+1}$$

$$c_i = 0, \quad i = 1, \ldots m, \quad c_{m+1} = 1$$

$$x^T = [w_1, w_2, \ldots, w_m, s]$$

In the dual case we choose the dual variable $Z \geq 0$ as

$$Z = \begin{bmatrix} z_1 \\ z_2 \end{bmatrix} \cdot [z_1^T \quad z_2^T] \tag{8}$$

where $z_1$, and $z_2$ are column vectors each with $|\mathcal{V}|$ elements. Obviously (8) choice of $Z$ implies that it is positive definite. From the complementary slackness condition (4) we have

$$(sI + W_1 - vv^T)z_1 = 0 \tag{9-a}$$

$$(sI - W_2)z_2 = 0 \tag{9-b}$$

Multiplying both sides of equation (9-a) by $vv^T$ we have $s(vv^T z_1) = 0$ which implies that

$$v^T z_1 = 0 \tag{10-a}$$

Using the constraints $Tr[F_i Z] = c_i$ we have

$$z_1^T z_1 + z_2^T z_2 = 1 \tag{11-a}$$

$$(\boldsymbol{\alpha}_i^T z_1)^2 = (\boldsymbol{\alpha}_i'^T z_2)^2 \quad \text{for} \quad i = 1, \dots, m \tag{11-b}$$

To have the strong duality we set $c^T x + Tr[F_0 Z] = 0$, hence we have

$$z_2^T z_2 - z_1^T z_1 = s \tag{12}$$

Considering the linear independence of $\boldsymbol{\alpha}_i$ and $\boldsymbol{\alpha}_i'$ for $i = 1, \dots, m$, we can expand $z_1$ and $z_2$ in terms of $\boldsymbol{\alpha}_i$ and $\boldsymbol{\alpha}_i'$ as

$$z_1 = \sum_{i=1}^{m} a_i \boldsymbol{\alpha}_i \tag{13-a}$$

$$z_2 = \sum_{i=1}^{m} a_i' \boldsymbol{\alpha}_i' \tag{13-b}$$

with the coordinates $a_i$ and $a_i'$, $i = 1, \dots, m$ to be determined.

Using (7) and the expansions (13), while considering (10), from comparing the coefficients of $\boldsymbol{\alpha}_i$ and $\boldsymbol{\alpha}_i'$ for $i = 1, \dots, m$ in the slackness conditions (9), we have

$$(-s - 1)a_i = -w_i \boldsymbol{\alpha}_i^T z_1, \tag{14-a}$$

$$(s - 1)a_i' = -w_i \boldsymbol{\alpha}_i'^T z_2, \tag{14-b}$$

where (14) holds for $i = 1, \dots, m$. Considering (11-b), we obtain

$$(s + 1)^2 a_i^2 = (s - 1)^2 a_i'^2, \tag{15}$$

for $i = 1, \dots, m$, or equivalently

$$\frac{a_i^2}{a_j^2} = \frac{a_i'^2}{a_j'^2} \tag{16}$$

for $\forall i, j = [1, m]$ and for $\boldsymbol{\alpha}_i^T z_1$ and $\boldsymbol{\alpha}_i'^T z_2$, we have

$$\boldsymbol{\alpha}_i^T z_1 = \sum_{j=1}^{n-1} a_j G_{i,j} \tag{17-a}$$

$$\boldsymbol{\alpha}_i^T z_2 = \sum_{j=1}^{n-1} a_j' G_{i,j}' \tag{17-b}$$

where $G$ and $G'$ are the Gram matrices, defined as

$$G_{i,j} = \boldsymbol{\alpha}_i^T \boldsymbol{\alpha}_j$$

$$G_{i,j}' = \boldsymbol{\alpha}_i^T \boldsymbol{\alpha}_j'$$

or equivallently

$$G = \begin{bmatrix} n+1 & -\sqrt{n} & 0 & \cdots & 0 \\ -\sqrt{n} & n+1 & -\sqrt{n} & 0 & \vdots \\ 0 & -\sqrt{n} & n+1 & \ddots & 0 \\ \vdots & 0 & \ddots & \ddots & -\sqrt{n} \\ 0 & \cdots & 0 & -\sqrt{n} & n+1 \end{bmatrix}$$

$$G' = \begin{bmatrix} 1 & \sqrt{n} & 0 & \cdots & 0 \\ \sqrt{n} & n+1 & -\sqrt{n} & 0 & \vdots \\ 0 & -\sqrt{n} & n+1 & \ddots & 0 \\ \vdots & 0 & \ddots & \ddots & -\sqrt{n} \\ 0 & \cdots & 0 & -\sqrt{n} & n+1 \end{bmatrix}$$

Substituting (17) in (14) we have

$$(s + 1 - (n+1)w_1)a_1 = -\sqrt{n}w_1 a_2 \tag{18-a}$$

$$(s + 1 - (n+1)w_i)a_i = -\sqrt{n}w_i(a_{i-1} + a_{i+1}) \quad \text{for} \quad i = 2, \ldots, m-1 \tag{18-b}$$

$$(s + 1 - (n+1)w_m)a_m = -\sqrt{n}w_m a_{m-1} \tag{18-c}$$

and

$$(-s + 1 - w_1)\hat{a}_1' = -\sqrt{n}w_1 a_2' \tag{19-a}$$

$$(-s + 1 - (n+1)w_2)a_2' = -\sqrt{n}w_2(\hat{a}_1' + a_3') \tag{19-b}$$

$$(-s + 1 - (n+1)w_i)a_i' = -\sqrt{n}w_i(a_{i-1}' + a_{i+1}') \quad \text{for} \quad i = 3, \ldots, m-1 \tag{19-c}$$

$$(-s + 1 - (n+1)w_m)a_m' = -\sqrt{n}w_m a_{m-1}' \tag{19-d}$$

where $\hat{a}_1' = -a_1'$. Now we can determine the optimal weights in an inductive manner as follows:

In the first stage, from comparing equations (18-c) and (19-d) and considering the relation (16), we can conclude that

$$(s + 1 - (n + 1)w_m)^2 = (-s + 1 - (n + 1)w_m)^2$$

which results in $w_m = 1/(n + 1)$ and $s = 0$, where the latter is not acceptable. Assuming $s = \frac{2\sqrt{n}}{n+1}\cos(\theta)$ and substituting $w_m = 1/(n + 1)$ in (18-c) and (19-d), we have

$$a_{m-1} = \frac{\sin(2(\pi - \theta))}{\sin(\pi - \theta)} a_m$$

$$a'_{m-1} = \frac{\sin(2\theta)}{\sin(\theta)} a'_m$$

Continuing the above procedure inductively, up to $i - 1$ stages, and assuming

$$a_j = \frac{\sin((m - j + 1)(\pi - \theta))}{\sin(\pi - \theta)} a_m, \qquad m > \forall j \geq i$$

and

$$a'_j = \frac{\sin((m - j + 1)\theta)}{\sin(\theta)} a'_m \qquad m > \forall j \geq i$$

for the $i$-th stage, by comparing equations (18-b) and (19-c) we get the following equations

$$\left((s + 1 - (n + 1)w_i)\frac{\sin((m - i + 1)(\pi - \theta))}{\sin(\pi - \theta)} + \sqrt{n}w_i \frac{\sin((m - i)(\pi - \theta))}{\sin(\pi - \theta)}\right) a_m = -\sqrt{n}w_i a_{i-1}, \qquad (20\text{-a})$$

$$\left((-s + 1 - (n + 1)w_i)\frac{\sin((m - i + 1)\theta)}{\sin(\theta)} + \sqrt{n}w_i \frac{\sin((m - i)\theta)}{\sin(\theta)}\right) a'_m = -\sqrt{n}w_i a'_{i-1} \qquad (20\text{-b})$$

and considering relation (16) we can conclude that

$$\left((s + 1 - (n + 1)w_i)\sin((m - i + 1)(\pi - \theta)) + \sqrt{n}w_i \sin((m - i)(\pi - \theta))\right)^2$$

$$= \left((-s + 1 - (n + 1)w_i)\sin((m - i + 1)\theta) + \sqrt{n}w_i \sin((m - i)\theta)\right)^2$$

which results in

$$w_i = \frac{1}{n + 1} \qquad (21)$$

Substituting $w_i = 1/(n + 1)$ in (20), we have

$$a_{i-1} = \frac{\sin\big((m-i+2)(\pi-\theta)\big)}{\sin(\pi-\theta)} a_m \tag{22-a}$$

$$a'_{i-1} = \frac{\sin\big((m-i+2)\theta\big)}{\sin(\theta)} a'_m \tag{22-b}$$

where (21) and (22) hold true for $i = 3, \ldots, m$ and in the $(m-1)$-th stage, from equations (18-b) and (19-b) and using relations (16) and (22), we can conclude that

$$w_2 = \frac{1}{n+1} \tag{23}$$

$$a_1 = \frac{\sin\big(m(\pi-\theta)\big)}{\sin(\pi-\theta)} a_m \tag{24-a}$$

$$\hat{a}'_1 = \frac{\sin(m\theta)}{\sin(\theta)} a'_m \tag{24-b}$$

and in the last stage, from equations (18-a) and (19-a) and using relations (16) and (22), we can conclude that

$$w_1 = \frac{2}{n+2} \tag{25}$$

and $\theta$ has to satisfy following relation

$$\sqrt{n}(n+1) \times \sin(m\theta) + n \times \sin\big((m-1)\theta\big) = (n+2)\sin\big((m+1)\theta\big) \tag{26}$$

where *SLEM* in terms of $\theta$ is

$$SLEM = \frac{2\sqrt{n}}{n+1} \cos(\theta). \tag{27}$$

Also one should notice that necessary and sufficient conditions for the convergence of weight matrix are satisfied, since all roots of $s$ which are the eigenvalues of $W$ are strictly less that one in magnitude, and one is a simple eigenvalue of $W$ associated with the eigenvector **1**, where this happens due to positivity of optimal weights [5].

Finally with the optimal weights given in (21), (23), (25), the dual constraints (11) and strong duality condition (12) are satisfied for $s = \frac{2\sqrt{n}}{n+1} \cos(\theta)$ with $\theta$ obtained from (26). Also $a_m^2$ and $a_m'^2$ determined as

$$a_m^2 = -\frac{\sin(\theta)}{2} \times \frac{\frac{2\sqrt{n}}{n+1}\cos(\theta) - 1}{\frac{2\sqrt{n}}{n+1}\cos(\theta) + 1} \times \frac{1}{f(\theta)}$$

$$a_m'^2 = -\frac{\sin(\theta)}{2} \times \frac{\frac{2\sqrt{n}}{n+1}\cos(\theta) + 1}{\frac{2\sqrt{n}}{n+1}\cos(\theta) - 1} \times \frac{1}{f(\theta)}$$

with

$$f(\theta) = \left(\frac{n}{(n+2)(n+1)}\right)\sin(\theta) + \frac{1}{n+1}\frac{\sin(m\theta/2)}{\sin(\theta/2)} \times \sin\left(\frac{(m+1)}{2}\theta\right).$$

## V. CONCLUSION

Fastest Distributed Consensus averaging Algorithm in sensor networks has received renewed interest recently, but Most of the methods proposed so far usually avoid the direct computation of optimal weights and deal with the Fastest Distributed Consensus problem by numerical convex optimization methods.

Here in this work, we have analytically solved fastest distributed consensus averaging problem for perfect $n$-ary tree network and complete $n$-ary tree network, by means of stratification and semidefinite programming. Our approach in this paper is based on convexity of fastest distributed consensus averaging problem, and inductive comparing of the characteristic polynomials initiated by slackness conditions in order to find the optimal weights. Also it has been indicated that the obtained optimal weights hold for perfect $n$-ary tree branch and complete $n$-ary tree branch independent of the rest of network.

We believe that the method used in this paper is powerful and lucid enough to be extended to networks with more general topologies which is the object of our future investigations.

**REFFERENCES**